\documentclass[12pt]{article}

\usepackage{graphicx}
 \usepackage{amssymb}
 \usepackage{amsmath}
 \usepackage{amsfonts}
 \usepackage{color}
 
\begin{document}

  \vspace{2cm}

  \begin{center}
    \font\titlerm=cmr10 scaled\magstep4
    \font\titlei=cmmi10 scaled\magstep4
    \font\titleis=cmmi7 scaled\magstep4
  {\bf The Casimir effect and mass renormalization \\
  for a  massive Bosonic string in  background B-field}

    \vspace{1.5cm}
    \noindent{{\large Y. Koohsarian\footnote{koohsarian\_ramian@yahoo.com}
   A. Shirzad\footnote{shirzad@ipm.ir}       }}\\
    \vspace{0.8cm}

   {\it Department of Physics, Isfahan University of Technology \\
P.O.Box 84156-83111, Isfahan, IRAN, \\
School of Physics, Institute for Research in Fundamental Sciences (IPM)\\
P.O.Box 19395-5531, Tehran, IRAN.}

  \end{center}

  \vskip 2em

\begin{abstract}
We study the Casimir effect for a massive bosonic string
terminating on D-brains, and living in a flat space with
an antisymmetric background B-field. We find the Casimir energy and
Casimir force as functions of the mass and length of the string
and show the force does not depend on B-field. We also find
the divergent part of the vacuum state energy, as a function
of the mass and length of the string and offer an idea for renormalization
of the mass of the string.
\end{abstract}


 \section{Introduction \label{sec1}}

Imposing definite  boundary conditions on a quantum field changes
the spectrum of the quantum states and leads in particular to
changing the vacuum energy of the system. These zero-point fluctuations
results in some observable quantum effects such as the well known
Casimir force \cite{Casimir}. As we know, this force depends on
the features of the space-time manifold and on the boundary
conditions imposed on the field.

A variety of theoretical models with different boundary conditions
have been considered in the literature in which the casimir effect
is analyzed and for some configurations the Casimir force is
observed or measured experimentally (see, e.g.,
\cite{Mos-Tur,Bord} as a review). The essential point for each
model is finding the Hamiltonian of the system as a combinations
of different physical modes (mostly harmonic oscillators) which
acquire positive excitation energies above the vacuum state. Then
turning off all of the excitations, one finds the vacuum state
energy of the system. The Casimier force emerges as the change in
the vacuum energy due to a small displacement of the boundaries.

In this paper we consider the model of a massive bosonic string in
a background B-field introduced initially in \cite{Ho-Cho1}. This
model is a generalization of the massless case which is a famous
model in the context of the string theory, specially because of
exhibiting noncommutative coordinates on the brains attached to
the endpoints of the string\cite{Ho-Cho2}. In a previous paper
\cite{Shir-Ba-Kooh}, considering the boundary conditions as Dirac
constraints and imposing them on the Fourier expansions of the
fields, we found the physical modes of the system as an infinite
set of harmonic oscillators. This enabled us to write down the
canonical Hamiltonian as a summation over Hamiltonians of simple
harmonic oscillators with definite frequencies. Hence, we can read
out the zero point energy as the summation over vacuum energy of
individual oscillators and regularize it to find out the Casimir
energy in terms of the length of the string. We will apply the
well known Abel-Plana formula for regularization of the vacuum
energy. Then by differentiating the Casimir energy with respect to
the string length, we will find the associated Casimir force as an
interaction between the D-brains (section 2).

Finally, we calculate the divergent part of the vacuum state
energy of the string in section 3. It's reasonable to interpret
the divergencies of the vacuum state energy as the renormalization
of some classical parameters of the system
\cite{Bord-Klim-Moh-Mos}. Here, we make a connection between the
divergent part of the vacuum state energy and renormalization of
the mass of the bosonic string.

\section{The Casimir force for the massive Bosonic string \label{sec2}}
Suppose an even number of fields, $X_i$, among Bosonic
fields $X^\mu$ living in a flat target space, are coupled to an
antisymmetric external tensor B-field. In the simplest case the
subspace of $X_i$'s is a two dimensional Euclidian space and the
constant B-field is exhibited by
\begin{equation}
B_{ij}\equiv \left( \begin{array}{cc} 0 & \tilde{B} \\
-\tilde{B} &0 \end{array} \right) . \label{e-1}
\end{equation}
Thus, neglecting those components of $X^\mu$ which does not couple
to the B-field, the simplified Lagrangian is given as
\cite{Ho-Cho1}
\begin{equation}
L=\frac{1}{2}\int_0^l d\sigma \left[\dot{X}^2-X'^2-m^2
X^2+2B_{ij}\dot{X}_iX'_j\right] \ ,\label{e-2}
\end{equation}
where "dot" and "prime" represent differentiation with respect to
$\tau$ and $\sigma$ respectively. Consistency of the variational
principal is achieved by considering the boundary conditions
${X_i}'+B_{ij}\dot{X}_j=0$ at the end-points $\sigma=0$ and
$\sigma=l$. In the canonical formulation the Hamiltonian reads
\begin{equation}
H=\frac{1}{2}\int_0^l d\sigma \left[(P-BX')^2+X'^2+m^2X^2\right] \
,\label{e-3}
\end{equation}
where $P_i=\dot{X}_i+ B_{ij}X'_j$ are conjugate momentum fields.
Hence, the boundary conditions can be considered as vanishing of
the primary constraint $\Phi_i(\sigma,\tau)= M_{ij}\partial_\sigma
X_j(\sigma,\tau)+B_{ij} P_j(\sigma,\tau)$ at the end-points
$\sigma=0$ and $\sigma=l$ where $M=1-B^2$. As shown in details in
\cite{Shir-Ba-Kooh}, the consistency of primary constraints, in
the language of constrained systems, gives the following two
infinite sets of constraints at the end-points
\begin{eqnarray}&&
(\partial_\sigma^2-m^2)^n\left[M_{ij}\partial_\sigma
X_j(\sigma,\tau)+B_{ij} P_j(\sigma,\tau)\right]=0, \nonumber \\
&& (\partial_\sigma^2-m^2)^n \left[\partial_\sigma
P_i(\sigma,\tau)-m^2B_{ij}X_j(\sigma,\tau)\right]=0, \label{e-4}
\end{eqnarray}
where $n=0,1,2,...$. Imposing the above constraints on the most
general Fourier expansions of the fields $X(\sigma,\tau)$ and
$P(\sigma,\tau)$, gives their expansions in terms of an enumerable
set of physical modes $a_n$ and $c_n$ as follow
\begin{eqnarray}
X(\sigma,\tau)=\frac{1}{\sqrt{l}} \left[a_0(\tau)\cosh {k_0
(\sigma -\frac{l}{2})}-\frac{1}{k_0}M^{-1}B
c_0(\tau)\sinh{k_0(\sigma-\frac{l}{2})}\right] \nonumber\\
+\sqrt{\frac{2}{l}}\sum_{n=1}^\infty \left[a_n(\tau)\cos
{\frac{n\pi}{l}\sigma}-\frac{l}{n\pi}M^{-1}Bc_n(\tau)\sin
{\frac{n\pi}{l}\sigma}\right], \nonumber \\
P(\sigma,\tau)=\frac{1}{\sqrt{l}}\left[c_0(\tau) \cosh{k_0(\sigma-
\frac{l}{2})}- \frac{1}{k_0}M^{-1}Ba_0(\tau)\sinh{k_0(\sigma-
\frac{l}{2})}\right] \nonumber\\
+\sqrt{\frac{2}{l}}\sum_{n=1}^\infty\left[c_n(\tau)\cos
{\frac{n\pi}{l}\sigma}-\frac{l}{n\pi}M^{-1}Ba_n(\tau)\sin
{\frac{n\pi}{l}\sigma}\right].\label{e-5}
\end{eqnarray}
Using the symplectic approach gives finally the classical brackets
of the physical modes as
\begin{equation}
[a_n , c_s]= N_n^{-1}\delta_{ns}, \label{e-6}
\end{equation}
where
\begin{equation}
N_0\equiv\frac{\sinh{k_0l}}{k_0l},\hspace{1cm} N_n\equiv1+
\frac{k_0^2l^2}{n^2\pi^2} \ \ n\neq0. \label{e-7}
\end{equation}
Inserting the expansions \eqref{e-5} of the fields in \eqref{e-3}
gives the Hamiltonian in terms of physical modes as
\begin{equation}
H=\frac{1}{2}\sum_{n=0}^\infty N_n(M^{-1}c_n^2+M\omega_n^2 a_n^2), \label{e-8}
\end{equation}
where
\begin{equation}
\omega_0^2=m^2M ,\hspace{1cm}
\omega_n^2=m^2+\frac{n^2\pi^2}{l^2} \hspace{5mm} n\ne 0. \label{e-9} \\
\end{equation}
The Hamiltonian \eqref{e-8} is, obviously, a superposition of
infinite number of independent harmonic oscillators with $a$'s as
positions and $c$'s as momenta.

Now, we can use these results to study the Casimir effect for the
current problem. From Eqs. \eqref{e-9} the zero-point energy of the system is
\begin{equation}
E_0(l,m)= \frac{1}{2}\left(\omega_0 + \sum_{n=1}^\infty
\omega_n\right)= \frac{1}{2}\left( m\sqrt{1+\tilde{B}^2} +
\sum_{n=1} ^\infty \sqrt{m^2+\frac{n^2\pi^2}{l^2}}\right) ,
\label{e-10}
\end{equation}
where we have used  the Planck units in which $\hbar=1$ and $c=1$.
The sum \eqref{e-10} is obviously infinite, as usual in quantum
field theory in assigning the ground state energy of a system. In
order to regularize \eqref{e-10}, we use a generalized form of the
known Abel-Plana formula \cite{Bord-Klim-Moh-Mos} as follow
\begin{equation}
\sum_{n=0}^\infty G_A (n) - \int_0^\infty dk G_A(k)
=\frac{1}{2}G_A(0) - 2\int_A^\infty \frac{dk}{exp(2\pi k)-1} (k^2
-A^2 )^\frac{1}{2} , \label{e-11}
\end{equation}
where $k$ is a continuous variable corresponding to $n$ and
$G_A(k)=\sqrt{A^2+k^2}$. To find the convergent part of
Eq. \eqref{e-10} we just have to take $G_m(n)=\frac{1}{2}
\sqrt{m^2+n^2\pi^2/l^2}$. After some simplifications we have
\begin{equation}
\frac{1}{2}\sum_{n=0}^\infty \sqrt{m^2+\frac{n^2\pi^2}{l^2}}
-\frac{1}{2}\int_0^\infty dk\sqrt{m^2+k^2}
=-\frac{m}{4}-\frac{1}{4\pi l}\int_\mu^\infty \frac{dy}{exp(
y)-1}\sqrt{y^2-\mu^2} , \label{e-12}
\end{equation}
where $y=2\pi k$ and $\mu=2ml$. So, comparing Eqs. \eqref{e-12}
and \eqref{e-10}  we obtain the Casimir energy as
\begin{equation}
E_c(l,m)=(\sqrt{1+\tilde{B}^2}-\frac{1}{2})\frac{m}{2} -
\frac{1}{4\pi l} \int_\mu^\infty
\frac{dy}{\exp(y)-1}\sqrt{y^2-\mu^2}. \label{e-13}
\end{equation}
Note that the result \eqref{e-13} has no dependence on the choice
of the regularization method.

Here, it seems necessary to attend to a physical points. If we
reasonably consider the parameter "$m$" as the mass of the bosonic
string, we anticipate the Casimir energy should vanish by taking
the limit $m\rightarrow\infty$. The reason is, for large values of
$m$, we expect the zero-point fluctuations of the vacuum state of
the string tend to zero. This requirement can be generally
considered as a universal condition for the Casimir energy of a
massive system \cite{Bord-Klim-Moh-Mos}. As a result of this
condition, the first term in Eq. \eqref{e-13} seems to be
problematic. However, this term may be absorbed as a
renormalization term for the mass of the string, as we will see in
the next section.

Now differentiating Eq. \eqref{e-13} with respect to $l$ gives the
Casimir force as follow
\begin{equation}
F_c(l,m)= -\frac{1}{4\pi l^2} \int_\mu ^\infty dy
\left(\frac{\mu^2}{[\exp(y) -1]\sqrt{y^2-\mu^2}}+\frac
{\sqrt{y^2-\mu^2}}{\exp(y)-1}\right). \label{e-14}
\end{equation}
For the massless bosonic string the corresponding results can be
obtained simply, by taking the limit $m\rightarrow 0$ in Eqs.
\eqref{e-13} and \eqref{e-14} as
\begin{eqnarray}
&&E_c(l)=-\frac{\pi}{24l} , \nonumber \\ && F_c(l)=-\frac{\pi}{24
l^2}. \label{e-15}
\end{eqnarray}
These results can also be obtained using the well known Zeta
function regularization.
\begin{figure}[!h]
\centering
\includegraphics[width=14cm,height=8cm]{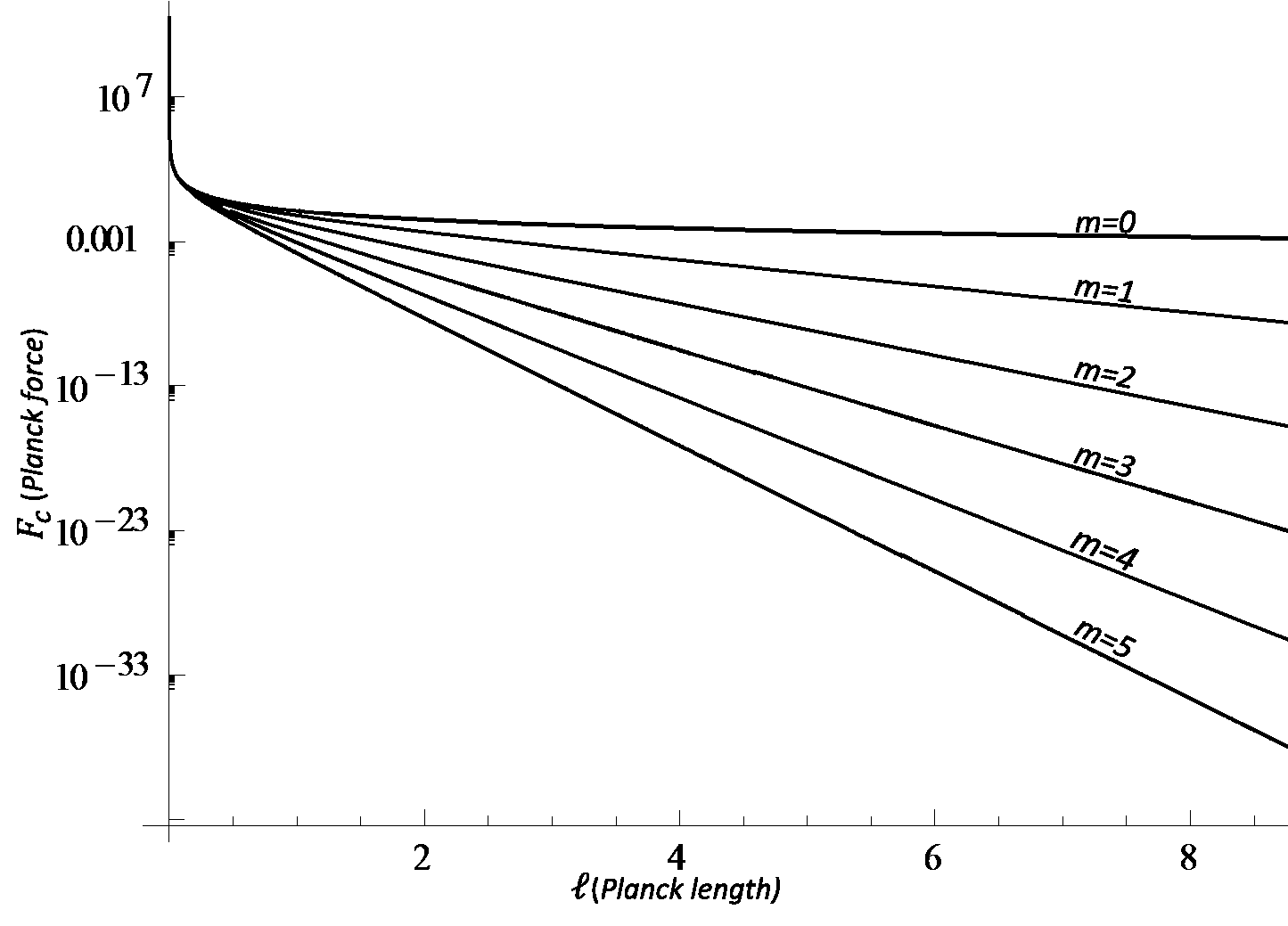}
\end{figure}
Obviously the Casimir force \eqref{e-15} associated with the
massless string as well as that of massive one \eqref{e-14}, has
no dependence on B-field. Hence, we conclude that the background
B-field does not play any role in the Casimir effect for massive
or massless bosonic string. as we see in above plot, the Casimir
force, in terms of Planck force ($F_p\approx 1.2 \times 10^{44}
N$), decreases when the string mass, in terms of Planck mass
($m_p\approx2.2\times10^{-8}kg$), or the string length, in terms
of Planck length ($l_p\approx 1.6\times 10^{-35}m$) increases, as
expected.

\section{ Mass Renormalization\label{sec3}}
In this section we want to show that the divergent part of the
zero-point energy should not be considered just as a redundant
part which should be canceled. As we will see, it may have, in
fact, physical interpretation.

The Abel-plana formula in the previous section just gave us
the convergent part of the vacuum energy.
To find the divergent part of the zero-point energy of the
string, we initially write
\begin{equation}
E_0(s)=\frac{\mu^{2s}}{2} \sum_{j=0}^\infty \omega_j^{1-2s},
\label{e-16}
\end{equation}
in which $\mu$  is a constant with the dimension of mass, included
for compensating the dimensions in opposite sides of Eq.
\eqref{e-16} and $j$ is the number of mode. Apparently $E_0(s)$ in
the limit $s\rightarrow0$ tends to vacuum state energy. From the
well-known definition of Gama function we can write
\begin{equation}
\omega_j^{1-2s}=\int_0^\infty\frac{dt}{t}\frac{t^{s-\frac{1}{2}}}
{\Gamma(s-\frac{1}{2})}\exp{(-t\omega_j^2)}. \label{e-17}
\end{equation}
For the massive bosonic string, using Eq. \eqref{e-9} we have
\begin{equation}
E_0(s)=\frac{\mu^{2s}}{2} \int_0^\infty\frac{dt}{t}\frac{t^{s-\frac{1}{2}}}
{\Gamma(s-\frac{1}{2})}\exp{(-t m^2)}\left(\sum_{n=1}^\infty\exp{
(-t\frac{n^2\pi^2}{l^2})}\right)+ \frac{\mu^{2s}}{2}\omega_0. \label{e-18}
\end{equation}
As is seen, the integrand in Eq. \eqref{e-18} is well-behaved for
large $t$. For small $t$, however, using the known Poisson
summation formula we obtain an asymptotic expansion as
\begin{equation}
\sum_{n=0}^\infty\exp{(-t\frac{n^2\pi^2}{l^2})}\approx
\frac{l}{\sqrt{4\pi t}}. \label{e-19}
\end{equation}
Hence, Eqs. \eqref{e-18} and \eqref{e-19} result in
\begin{eqnarray}
E_0(s)\approx\frac{\mu^{2s}}{2}\frac{l}{\sqrt{4\pi}} \int_0^\infty
\frac{dt}{t}\frac{t^{s-1}}{\Gamma(s-\frac{1}{2})}\exp{(-t m^2)}+
\frac{\mu^{2s}}{2}\omega_0 \nonumber \\ =\frac{\mu^{2s}}{2}\frac{lm^2}
{\sqrt{4\pi}}\frac{\Gamma(s-1)}{\Gamma(s-\frac{1}{2})}+
\frac{\mu^{2s}}{2}\omega_0. \label{e-20}
\end{eqnarray}
Here, an allusion to the theme of the familiar expansion of heat
kernel may seem interesting. It is well-known that the heat kernel
$K(t)$, has an asymptotic expansion for $t\approx0$ as follows
\begin{equation}
K(t)\approx\frac{1}{\sqrt{(4\pi t)^d}}\sum_{n=0}^\infty a_nt^n, \label{e-21}
\end{equation}
in which, $d$ is the dimension of the space where the dynamical
fields live in, and $a_n$'s are recognized as the heat kernel
coefficients. Comparing Eqs. \eqref{e-19} and \eqref{e-21}, we
deduce that the heat kernel asymptotic expansion for the bosonic
string can be derived, simply, if we take $d=1$, $a_0=l$ and
$a_i\approx0$, for $i \ge 1$ . In other word, since $a_0$ is often
considered as the volume of the underlying space, the bosonic
string, in the framework of heat kernel expansion, can be seen
as a one dimensional system with volume $l$.
Now turning back to Eq. \eqref{e-20}, we have for $s\rightarrow0$
\begin{equation}
 E_0(s)\approx\frac{lm^2}{8\pi}\frac{1}{s}+\frac{\omega_0}{2}, \label{e-22}
\end{equation}
where the asymptotic value $\Gamma(s-1)\approx-\frac{1}{s}$ for
$s\approx0$, has been employed and also we have inserted
$\Gamma(-\frac{1}{2})=-2\sqrt{\pi}$. So the divergent part of the
vacuum state energy for the massive bosonic string is
\begin{equation}
E_0^{\textrm{div}}(s)=\frac{lm^2}{8\pi}\frac{1}{s},\hspace{0.5cm}
s\rightarrow0. \label{e-23}
\end{equation}
Note that $E_0^{\textrm{div}}(s)$ in Eq.\eqref{e-23} is
proportional to the mass squared as well as the length of the
string; hence, it may have some physical significance. Using the
idea given in Ref. \cite{Bord-Klim-Moh-Mos}, it would be more
clear if we reasonably, add the vacuum state energy to the
classical ground state energy $E_{0,class}$ , to evaluate the
total ground state energy, $E_{0,total}$, of a system, as follows
\begin{eqnarray}
E_{0,total}=E_{0,class}+E_0(s) \nonumber
=\underbrace{E_{0,class}+E_0^{div}(s)}+\underbrace{E_0(s)-E_0^{div}(s)} \nonumber\\
=\hspace{1.5cm}E_{0,class}^{ren}\hspace{1cm}+\hspace{1cm}E_{0,qu}.
\label{e-24}
\end{eqnarray}
Here we have subtracted the divergent part from vacuum state
energy and appended it to the classical ground state energy, as a
reasonable interpretation for the renormalization of the classical
ground state energy of the system.

For the relativistic massive string, it is reasonable to consider
the mass of the string, as the classical ground state energy,
$E_{0,class}=m$. Hence we can write
\begin{equation}
E_{0,class}^{ren}=m+\frac{lm^2}{8\pi}\frac{1}{s},\hspace{0.3cm}
s\rightarrow0.\label{e-25}
\end{equation}
Eventually, we can think of Eq.\eqref{e-25}, as renormalization of the
mass of bosonic string as
\begin{equation}
m\rightarrow m+\frac{lm^2}{8\pi}
\frac{1}{s},\hspace{0.3cm} s\rightarrow0, \label{e-26}
\end{equation}
As we said in the previous section, we should impose the physical
requirement $E_{0,qu}\rightarrow0\hspace{0.3cm} for\hspace{0.2cm}
m\rightarrow\infty$. To fulfill this condition the first term in
Eq. \eqref{e-13} can be considered as another mass renormalization
term. Hence, the final expression for renormalization of the
string mass can be written as follow
\begin{equation}
 m\rightarrow m+(\sqrt{1+\tilde{B}^2}-\frac{1}{2})\frac{m}{2}+\frac{lm^2}{8\pi}
\frac{1}{s},\hspace{0.3cm} s\rightarrow0. \label{e-27}
\end{equation}
Note, the last two terms in Eq. \eqref{e-27} are resulted from
quantization of the system, and vanish in the classical limit
$\hbar\rightarrow0$. We see that the background B-field plays a
role in the mass renormalizaton, although it has no influence on
the Casimir effect for the Bosonic string, as we previously
realized.


\begin{thebibliography}{99}

\bibitem{Casimir}H. B. G. Casimir, Proc. K. Ned. Akad. Wet 51, 793 (1948).

\bibitem{Mos-Tur}V. M. Mostepanenko and N.N. Turnov,
{\it{The Casimir effect and its applications}},
(Oxford University Press, Oxford, 1997).

\bibitem{Bord}M. Bordag, {\it The Casimir Effect. 50 years
latre} (World Scientific, Singapore, 1999).

\bibitem{Ho-Cho1}C.-s. Cho and P.-M. Ho,
{\it{ Non-commutative  D-brane  and  open  string  in  pp-wave
 background  with B-field}} Nucl. Phys. B \textbf{636}, 141-158, (2002).

\bibitem{Ho-Cho2}C.-s. Cho and P.-M. Ho, {\it{Noncommutative
open  string  and  D-brane}} Nucl. Phys. B \textbf{550},
151-168, (1999), hep-th/9812219.

\bibitem{Shir-Ba-Kooh} A. Shirzad, A. Bakhshi and Y. Koohsarian,
hep-th 1112.5781

\bibitem{Bord-Klim-Moh-Mos} M. Bordag, G. L. Klimchitskaya, U.
Mohideen and V. M. Mostepanenko, {\it{ Advances in Casimir
Effect}}, (Oxford Science Publications 2008).

\end{thebibliography}
\end{document}